\documentclass[aps,prl,showpacs,twocolumn]{revtex4}
\usepackage{amssymb}
\usepackage{epsfig}
\usepackage{graphicx}
\usepackage{amsmath}
\usepackage{array,color}

\begin{document}

\title{Effect on Cavity Optomechanics of the Interaction
Between a Cavity Field and a 1D Interacting Bosonic Gas }
\author{Qing Sun$^{1,2}$, Xing-Hua Hu$^1$, W. M. Liu$^1$, X. C. Xie$^{3,4}$,
and An-Chun Ji$^{2,1}$}

\address{$^1$Beijing
National Laboratory for Condensed Matter Physics, Institute of
Physics, Chinese Academy of Sciences, Beijing 100190, China}

\address{$^2$Department of Physics,
Capital Normal University, Beijing 100048, China}

\address{$^3$International Center for Quantum Materials, Peking
University, Beijing 100871, China}

\address{$^4$Department of Physics, Oklahoma State University,
Stillwater, Oklahoma 74078, USA}

\date{{\small \today}}

\begin{abstract}
We investigate the optomechanical coupling between 1D interacting
bosons and the electromagnetic field in a  high-finesse optical
cavity. We show that by tuning the interatomic interactions, one can
realize the effective optomechanics  with the mechanical resonators
ranging from the side-mode excitations of a Bose-Einstein condensate
(BEC) to particle-hole excitations of Tonks-Girardeau (TG) gas. We
propose that, this unique feature can be formulated to detect the
BEC-TG gas crossover and measure the sine-Gordon transition
continuously and nondestructively, which are achievable immediately
in current experiments.
\end{abstract}
 \maketitle

The experimental achievements in manipulating the strong coupling
between ultracold atoms and the electromagnetic field in an optical
cavity have triggered many new exciting advances to cavity quantum
electrodynamics (QED) \cite{Brennecke1,Colombe,Baumann,Gupta,Murch,
Nagy,Keeling,Brennecke2}. One of the remarkable achievements is to
implement cavity optomechanics with cold atoms \cite{Gupta,Murch} or
a BEC \cite{Brennecke2}, which is of great importance both for
technical applications ranging from optical communication to quantum
computation \cite {Kippenberg}, and conceptional exploration of the
classic-quantum boundaries \cite{Braginsky}.

In this Letter, we investigate the optomechanical coupling between a
1D  interacting bosonic gas and a cavity field. So far the recent
works have neglected the interatomic interactions or considered
merely the weakly interacting region, where the mean-field
Bogoliubov theory is valid \cite{Brennecke2,Goldbaum}. In this case,
the 1D bosonic gas forms a BEC (or quasi-condensate) \cite{Gorlitz},
and the side-mode excitations of the condensate play the role of
mechanical resonator with the bare frequency $\omega_M^{0}=4\hbar
k^2/M$ \cite{Brennecke2}, here $k=2\pi/\lambda_c$ is the wave-vector
of the cavity mode. However, when the interatomic interactions are
added into the system, the situation changes dramatically. The
strong interatomic interactions would transform the ground state of
the condensate to a Luttinger liquid (LL); and remarkably in the
strongly interacting limit, the 1D bosons--known as a TG gas
\cite{Girardeau,Lieb,Paredes,Kinoshita}--exhibit completely
different behavior like ideal fermions. It is therefore important to
explore the interatomic interaction effects on the cavity
optomechanics, where the quantum fluctuations of 1D bosons are very
strong.

In this work, we first employ the quantum hydrodynamical approach to
derive an effective model of the cavity QED with 1D interacting
bosons. We show that  the effective optomechanics can be realized in
the intermediately and strongly interacting regions. The
corresponding optomechanical coupling is determined by the
low-energy excitations or the interatomic interactions of the
bosons. Therefore, by probing the cavity oscillations or the noise
spectra versus the interatomic interactions, one can determine the
quantum phases of the 1D interacting gas and detect the BEC to TG
gas crossover, a fascinating phenomenon of the system\cite{Moritz}.
Furthermore, we propose that one could also measure the sine-Gordon
transition, which has stimulated considerable interest
\cite{Buchler,Haller}, conveniently with the nondemolition
measurements \cite{nondemolition,Mekhov}.

\begin{figure}[t]
\centering
\includegraphics[width=0.4\textwidth]{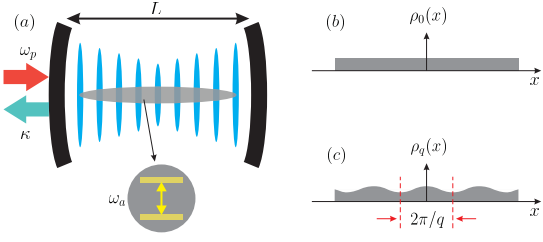}
\caption{(color online). Experimental set-up and schematic density
distribution.  (a) $N$ bosonic atoms with resonant frequency
$\omega_a$ are confined in an effectively 1D trap inside an optical
cavity of length $L$. The cavity mode is driven by a pump laser of
frequency $\omega_p$, and $\kappa$ is the decay rate. (b)
Ground-state atomic density distribution $\rho_0(x)$ in the absence
of a cavity mode. (c) Distribution of density fluctuation
$\rho_q(x)$ with wave-vector $q=\pm2k$, which is scattered by the
periodic potential of the cavity mode.}\label{set-up}
\end{figure}

The system under investigation is schematically depicted in Fig.
\ref{set-up}(a), where $N$ ultracold bosonic atoms of mass $M$ with
resonant frequency $\omega_a$ are confined in a 1D trap inside an
optical cavity with length $L$. The cavity mode of frequency
$\omega_c$ is driven by a pump laser of frequency $\omega_p$ at rate
$\eta$, and $\kappa$ is the decay rate of the cavity field.
Following Ref. \cite{Maschler}, we adiabatically eliminate the
internal excited state of the atoms, as justified by the large
detuning between the atomic resonance and pump frequency. Then, by
using the dipole and rotating-wave approximations, one arrives at
the following Hamiltonian of the atomic part
\begin{eqnarray}
\hat{H}_a&=&\frac{\hbar^2}{2M}\int^{L}_{0}
dx\partial_x\hat{\Psi}^{\dag}(x)\partial_x\hat{\Psi}(x)+\int^{L}_{0}
dx\hat{V}(x)\hat{\rho}(x)\nonumber\\
& &+\>\frac{1}{2}\int^{L}_{0}
dxdx^{\prime}\hat{\rho}(x)U(x-x^{\prime})\hat{\rho}(x^{\prime}).
\label{Hamiltonian1}
\end{eqnarray}
Here, $\hat{\Psi}(x)$ is the bosonic field operator and
$\hat{V}(x)=\hbar U_0\cos^2(kx)\hat{c}^{\dag}\hat{c}$  is  the
dynamical periodic potential, with $\hat{c}$ the annihilation
operator of a cavity photon, and the potential depth
$U_0=g_0^2/(\omega_p-\omega_a)$. The interatomic interactions are
given by contact pseudo-potentials
$U(x-x^{\prime})=g_{1d}\delta(x-x^{\prime})$, where
$g_{1d}=\frac{2\hbar^2a_s}{(1-\mathcal{C}a_s/\sqrt{2}l_{\perp})
Ml^2_{\perp}}$ is the effective 1D coupling strength with $a_s$  the
three-dimensional scattering length, $\mathcal{C}=1.0325$, and
$l_{\perp}=\sqrt{\hbar/M\omega_{\perp}}$ the transverse oscillator
length.

We start by considering the general situation with arbitrary
interatomic interactions and derive an effective model of the the
system by using quantum hydrodynamical approach \cite{Haldane},
which is a well-defined low-energy theory. We shall work in the low
photon numbers limit, where the dynamical periodic potential
$\hat{V}(x)$ is feeble. By introducing two new fields
$\hat{\phi}(x)$ and $\hat{\theta}(x)$, which describe the collective
fluctuations of the phase and density respectively and satisfy the
commutation relation
$[\hat{\phi}(x),\frac{1}{\pi}\partial_{x^{\prime}}\hat{\theta}
(x^{\prime})]=i\delta(x-x^{\prime})$, we can express the bosonic
field operator $\hat{\Psi}(x)$ as
\begin{eqnarray}
\hat{\Psi}(x)\sim[\rho_0-\frac{1}{\pi}\partial_x\hat{\theta}(x)]^{1/2}
\{\sum^{+\infty}_{m=-\infty}e^{2mi\hat{\theta}(x)}e^{i\hat{\phi}(x)}\},
\label{field}
\end{eqnarray}
and the corresponding density operator
\begin{eqnarray}
\hat{\rho}(x)=[\rho_0-\frac{1}{\pi}\partial_x\hat{\theta}(x)]
\sum^{+\infty}_{m=-\infty}e^{2im(\hat{\theta}(x)-\pi\rho_0x)}.
\label{density}
\end{eqnarray}

In the following, we shall first consider the long-wavelength
approximation, i.e. $\lambda_c\gg1/\rho_0$, where we can only keep
the $m=0$ term in Eqs. (\ref{field}) and (\ref{density}). In this
limit, the system can be expressed by the hydrodynamical
description, with the weak dynamical periodic potential coupled to
the slow part of the density operator $\hat{\rho}(x)$. Then, the
corresponding low-energy effective Hamiltonian of the atomic part
reads
\begin{eqnarray}
\hat{H}^{\prime}_a&=&\int^{L}_{0}dx\{\frac{\hbar\upsilon_s}{2\pi}
[K(\partial_x\hat{\phi}(x))^2+
\frac{1}{K}(\partial_x\hat{\theta}(x)-\pi\rho_0)^2]\nonumber\\
&&-\frac{\hat{V}(x)}{\pi}\partial_x\hat{\theta}(x)\}.
\label{Hamiltonian2}
\end{eqnarray}
Here, $K$ is the dimensionless parameter  and $\upsilon_s$  the
sound velocity. They both depend on a single dimensionless
interacting parameter $\gamma=Mg_{1d}/\hbar^2\rho_0$,  with
$\upsilon_sK\equiv\upsilon_F=\hbar\pi\rho_0/M$ fixed by Galilean
invariance. We note that, the Hamiltonian (\ref{Hamiltonian2})
describes a LL coupled to a weak periodic potential, which is
dynamically dependent on the atomic state and determined
self-consistently.

We further transform the Hamiltonian (\ref{Hamiltonian2}) to
momentum representation and then implement the standard bosonization
procedure by introducing the bosonic creation operator
$\hat{b}^{\dag}_q=\sqrt{\frac{2\pi}{|q|L}} \hat{\rho}_q$. We can
arrive at the following effective Hamiltonian of the coupled system
\begin{eqnarray}
\hat{H}_{\>\rm
eff}&=&\sum_{q=\pm2k}\hbar\omega_q\hat{b}^{\dag}_q\hat{b}_q +\hbar
g\sum_{q=\pm
2k}(\hat{b}^{\dag}_q+\hat{b}_q)\hat{c}^{\dag}\hat{c}+\hbar\Delta
\hat{c}^{\dag}\hat{c}\nonumber\\
&&+\>i\hbar\eta(\hat{c}^{\dag}-\hat{c}). \label{Hamiltonian3}
\end{eqnarray}
Here the first term describes the long-wavelength
density-fluctuations of the 1D interacting gas with
$\omega_q=\upsilon_s|q|$ for $|q|\ll\rho^{-1}_0$. The second term is
the coupling between the corresponding density-fluctuations and
cavity field with $g=\frac{U_0}{4}\sqrt{\frac{kL}{\pi}}$. In the
sums, we assume that only the $q=\pm2k$ modes are coupled to the
cavity, which is justified by the low photon numbers limit.
$\Delta=\omega_c-\omega_p+U_0N/2$ is the effective cavity detuning.

The above effective Hamiltonian (\ref{Hamiltonian3}) actually
describes the optomechanical coupling between a mechanical
oscillator (with the frequency $\omega_M=\omega_{\pm
2k}=2k\upsilon_s$) and the radiation pressure force of a cavity
field. To see this, we introduce the quadratures of the  bosonic
excitations
$\hat{X}_M=\sum_{q=\pm2k}(\hat{b}^{\dag}_q+\hat{b}_q)/\sqrt{2}$, and
derive the following Heisenberg-Langevin equations
\begin{eqnarray}
&&\frac{d^2\hat{X}_M}{dt^2}+\omega^2_M\hat{X}_M=-2\sqrt{2}
g\omega_M\hat{c}^{\dag}\hat{c},
\label{Dynamics1}\\
&&\frac{d\hat{c}}{dt}=-i\Delta_{\>\rm eff}\>\hat{c}+\eta
-\kappa\hat{c}+\sqrt{2\kappa}\hat{c}_{\rm\> in}, \label{Dynamics2}
\end{eqnarray}
with the resonance frequency $\Delta_{\>\rm
eff}=\Delta+\sqrt{2}g\hat{X}_M$ and $\sqrt{2\kappa}\hat{c}_{\rm\>
in}$ the noise term.  Here, the low-energy long-wavelength density
fluctuation (phonon) plays the role of a mechanical resonator.

\begin{figure}[t]
\centering
\includegraphics[width=0.47\textwidth]{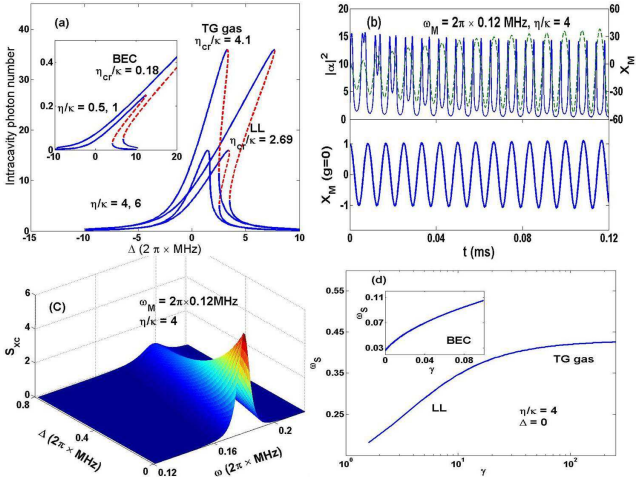}
\caption{(color online). Steady-state and dynamical behavior of the
optomechanical coupling. (a) Mean cavity photon number versus
$\Delta$  for $\omega_M=2\pi\times [0.02\>({\rm BEC}), 0.12\>({\rm
LL}), 0.28\>({\rm TG\>gas})]$ MHz. Here $\eta_{\rm cr}$ is the
bistability threshold. (b) Cavity photon number $|\alpha|^2$,
coupled oscillator $X_M$ (dashed line), and free oscillator
$X_M(g=0)$ for $\Delta=0$. (c) Noise spectrum $S_{\rm X_c}$ versus
$\omega$ and $\Delta$. (d) Center frequency $\omega_S$ of $S_{\rm
X_c}$ versus $\gamma$. Inset shows the results for small $\gamma$
region.}\label{solution}
\end{figure}

Now, we investigate the optomechanical coupling governed by the set
of coupled Eqs. (\ref{Dynamics1})-(\ref{Dynamics2}).  One of the
characteristic phenomenon of cavity optomechanics is the bistable
behavior of the stationary solutions, which we derive as
$\bar{X}_M=-2\sqrt{2}g|\bar{c}|^2/\omega_M$, and $
|\bar{c}|^2=\eta^2/[\kappa^2+(\Delta-4g^2\omega^{-1}_M|
\bar{c}|^2)^2]$. In Fig. \ref{solution}a, we give the bistability
for three typical oscillator frequency $\omega_M$, which correspond
to different quantum phases of the system (see below). The linear
stability analysis shows that the middle branch (dashed line) is
unstable, while the up and low branches are stable. Here, the
typical experimental parameters are used: $L\sim100$ $\mu$m,
$\lambda_c=780$ nm, $N\simeq 5000$ $^{87}$Rb atoms, with
$1/\rho_0\simeq20$ nm$\ll\lambda_c$ satisfing the wave-length
approximation; $\kappa=2\pi\times 1$ MHz, $U_0=2\pi\times 20$ kHz
with the Rabi frequency $g_0=2\pi\times10.9$ MHz and the pump-atom
detuning $\omega_p-\omega_a=2\pi\times32$ GHz.

To discuss the dynamics of the optomechanics, we have taken into
account the lossy and driven cavity, where quantum jumps in the
cavity photon number can lead to a strong entanglement between the
cavity photon number and bosonic wavefunction. This creates a
displacement noise spectrum of the mechanical oscillator
$S_{X_M}(\omega)=2\kappa(4g\bar{c}\omega_M)^2[\kappa^2
+(\tilde{\Delta}+\omega)^2]/|d(\omega)|^2$, and the corresponding
measurable noise spectrum of the cavity field quadrature
$\hat{X}_c=(\hat{c}^\dagger+\hat{c})/\sqrt{2}$ \cite{Paternostro},
which gives $S_{X_c}(\omega)=[(2g\bar{c}\tilde{\Delta})^2
S_{X_M}(\omega)+2\kappa[\kappa^2+
(\tilde{\Delta}+\omega)^2]]/|d(\omega)|^2$ (see Fig.
\ref{solution}c). Here,
$d(\omega)=(\omega^2-\omega_M^2)[(\kappa-i\omega)^2+\tilde{\Delta}^2]
+2\omega_M\tilde{\Delta}(2g\bar{c})^2$, and
$\tilde{\Delta}=\Delta-(2g\bar{c})^2/\omega_M$.

\begin{figure}[t]
\centering
\includegraphics[width=0.4\textwidth]{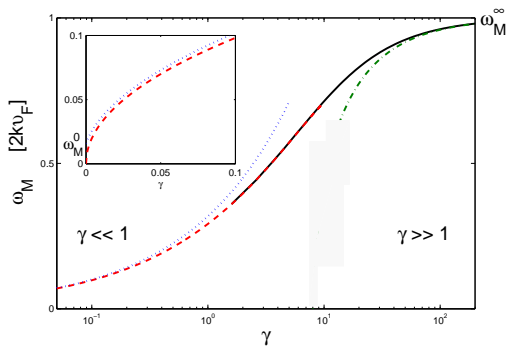}
\caption{(color online). Mechanical oscillator frequency $\omega_M$
versus $\gamma$. The solid line is the numerical Bethe-Ansatz
result. The dashed and the dash-dotted lines are the asymptotic
results in the weakly and strongly interacting regions respectively.
In the weakly interacting region, we also give the Bogoliubov
excitation spectrum (the dotted line).}\label{frequency}
\end{figure}

We then explore the interatomic interactions dependence of the
effective optomechanics. First we note that, whether the system is
in weakly or strongly interacting regions, one can realize the
effective mechanical oscillator  in the whole regions. This implies
that the 1D bosonic gas is in fact in a universal class, which can
be well described by the low-energy hydrodynamical theory. However,
the mechanical oscillator frequency $\omega_M$ is fully dependent on
interatomic interaction parameter $\gamma$  with
$\omega_M=2k\upsilon_s$. Here, the  sound velocity is given by
$\upsilon_s=\sqrt{(\rho_0/M)\>\partial^2E/\partial N^2}$, and
$E=\sum_l \hbar^2k^2_l/2M$ is the ground-state energy of 1D bosonic
gas with $k_l$ determined by the following Bethe-Ansatz equations
\cite{Lieb}
\begin{eqnarray}
k_l L&=& 2\pi I_l-\sum_{m=1}^N
\tan^{-1}\left(\frac{k_l-k_m}{\gamma\rho_0}\right),
\label{Bethe-Ansatz}
\end{eqnarray}
where $I_l\in \{-(N-1)/2, \cdots, (N-1)/2\}$ are the set of
integers. We solved Eq. (\ref{Bethe-Ansatz}), and in Fig.
\ref{frequency} we show  the numerical results of $\omega_M$ versus
$\gamma$. Experimentally, when  photons enter the cavity, the light
filed and the bosonic excitations are coupled nonlinearly, the
eigenfrequency of Eqs. (6)-(7) will modify $\omega_M$. Nevertheless,
the optomechanical coupling is fully determined by the frequency
$\omega_M$ of bosonic excitations and thereby dependent on $\gamma$
of the 1D gas. Accordingly, by measuring on the cavity field
oscillations or detecting the noise spectrum, one can determine the
continuous BEC-TG gas crossover, which is an intriguing result of
the system.

In the weakly interacting region, we can use the  Bogoliubov
approximation to derive
$\omega_M=2k\upsilon_F\sqrt{\gamma-\gamma^{3/2}/(2\pi)}/\pi$ for
$\gamma\leq 10$ (the dashed line of Fig. \ref{frequency}). However,
we note that $\omega_M$ vanishes as $\gamma\rightarrow0$,  which
contradicts with recent experiments \cite{Brennecke2}. In fact, as
we know because the bosons begin to (quasi) condense  for $\gamma\ll
1$, the dominant contribution of the density fluctuation
$\hat{\rho}_{\pm2k}=\hat{\Psi}^{\dag}_{\pm2k}
\hat{\Psi}_{q=0}+\sum_{q\neq0}
\hat{\Psi}^{\dag}_{\pm2k+q}\hat{\Psi}_{q}$ is the quasi-particle
excitations from macroscopic occupied $q=0$ ground state. The energy
of a quasiparticle  is then determined by the Bogoliubov excitation
spectrum
$\omega_M=\sqrt{\epsilon_{\pm2k}(\epsilon_{\pm2k}+2g_{1d}\rho_0)}/\hbar$
(the dotted line  of Fig. \ref{frequency}). In this case, the
mechanical oscillators are the side-mode excitations of a BEC; and
in the limit of $\gamma=0$ we get the bare oscillator frequency
$\omega_M^{0}=4\hbar k^2/M$ (inset of Fig.\ref{frequency}).
Accordingly, the coupling between the oscillator and a cavity field
should be replaced by $g=U_0/2 \sqrt{N/2}$ in Eqs.
(\ref{Dynamics1})-(\ref{Dynamics2}), which is enhanced by the
condensation. We give the stationary bistability in the inset of
Fig. \ref{solution}a, and then numerically integrate the coupled
equations for $\Delta=0$ by switching on $\eta/\kappa=4$ at $t=0$.
We find that the cavity field oscillates regularly (not shown), but
the frequency of the cavity field has a large shift of $\omega_M$,
which agrees well with Ref. \cite{Brennecke2}.

When $\gamma$ is further increased above unity, the collective
density excitations become dominant and the 1D gas crosses to a LL
\cite{Moritz}. Numerical integration of Eqs.
(\ref{Dynamics1})-(\ref{Dynamics2}) shows that both $|\alpha|^2$ and
$X_M$ exhibit well-defined oscillations, and  the cavity field is
excited resonantly at $X_M=0$ (Fig. \ref{solution}b). Yet we note
that, different from the BEC phase where $g$ is collectively
enhanced, the optomechanical coupling becomes small. In this case,
the linear stability analysis shows that the eigenfrequency of the
set of coupled equations is nearly the same with the free oscillator
$\omega_M$, see Fig. \ref{solution}b for example. This is a
characteristic phenomenon of the region. Experimentally, by
increasing the interatomic interactions, if the oscillation
frequency of the cavity field follows the the solid line of Fig.
\ref{frequency}, the 1D gas should be in the LL phase. We also
calculate the noise spectrum $S_{\rm X_c}$ (Fig. \ref{solution}c),
where the center frequency $\omega_S$ of the spectrum has a shift of
$\omega_M$. In Fig. \ref{solution}d, we give $\omega_S$ for
$\Delta=0$ in the whole interacting regions. We find that $\omega_S$
increases with $\gamma$ and has a tendency to saturate above
$\gamma\simeq 50$, which can be inspected in experiments.

While for $\gamma\gg 1$, the strong interactions would prevent the
bosons from occupying the same position. Especially when
$\gamma=\infty$, the symmetric many-body wave function of bosons can
be  mapped to an antisymmetric fermionic wave function by
$\Psi_B(x_1,\cdots,x_N)=A(x_1,\cdots,x_N)\Psi_F(x_1,\cdots,x_N)$
with $A(x_1,\cdots,x_N)=\prod_{1\leq j<k\leq N}{\rm sgn}(x_k-x_j)$
\cite{Girardeau}. Hence, the Hamiltonian (\ref{Hamiltonian1}) can be
rewritten in terms of the fermion field operators
\begin{eqnarray}
\hat{H}_{F}\!\!&=&\!\!\frac{\hbar^2}{2M}\!\!\int^{L}_{0}\!\!\!\!
dx\partial_x\hat{\Psi}^{\dag}_{F}(x)\partial_x\hat{\Psi}_{F}(x)
\!+\!\!\!\int^{L}_{0}\!\!\!\!
dx\hat{V}(x)\hat{\rho}_{F}(x),\label{Hamiltonian4}
\end{eqnarray}
which is exactly the model describing a free  fermion gas subjecting
to the cavity periodic potential \cite{Kanamoto}. Then, the
oscillator is formed by the particle-hole excitations at the edges
of $\pm k_F$ through the Bose-Fermi mapping, and the mechanical
frequency is naturally related to the Fermi velocity:
$\omega_M^\infty=2k\upsilon_F=2k\hbar\pi\rho_0/M$. While for finite
$\gamma$, we use the asymptotic expression to derive $\omega_M=
2k\upsilon_F[1-4/\gamma]$ (the dash-dotted line of Fig.
\ref{frequency}). Experimentally, when the 1D gas becomes a TG gas,
the oscillation frequency of the cavity field should follow this
asymptotic expression and approaches $\omega_M^\infty$.
Correspondingly, the center frequency $\omega_S$ will saturate at
$2\pi\times 0.42$ MHz (Fig. \ref{solution}d).

Finally, let us turn to the commensurate situation with
$\lambda_c\sim2/\rho_0$, where a new instability--the sine-Gordon
transition--may appears in the strongly interacting 1D quantum gas
\cite{Haller}: the superfluid ground state turns insulating in the
presence of a weak commensurate periodic potential. It is now
necessary to take account of the discrete nature of boson with
$m\neq0$ terms in Eq. (\ref{density}). This gives rise to a
sine-Gordon type perturbation \cite{Buchler} up to the leading term
\begin{eqnarray}
\hat{H}_{\rm s-G}=\frac{1}{2}\hbar
U_0\hat{c}^{\dag}\hat{c}\rho_0\int^{L}_{0}dx\cos[2\hat{\theta}(x)+Qx],
\end{eqnarray}
where $Q=2\pi(\rho_0-k/\pi)$, which vanishes at commensurability. In
the small photon numbers limit, it was shown that \cite{Gogolin},
this term is renormalization irrelevant for $K>K_c=2$ or
equivalently $\gamma<\gamma_c=3.5$, leaving the ground state a
superfluid LL with the same linear excitation spectrum as the
long-wavelength approximation. Then, one expects a well-defined
oscillation of optomechanical coupling. While for $\gamma>\gamma_c$,
$\hat{H}_{\rm s-G}$ becomes relevant, the system transits to an
insulating Mott phase, where the bosonic excitations are forbidden
owing to the energy cost, and the optomechanical oscillation
vanishes correspondingly. Therefore, we can  use the optomechanical
dynamics across the critical point $\gamma_c$ to detect the
sine-Gordon transition easily in experiments.

In summary, we demonstrate that one can realize the effective
optomechanics in the whole interacting regions of 1D bosonic gas.
This offers  a new approach to to detect the BEC-TG gas crossover or
the sine-Gordon transition by investigating the optomechanical
coupling. These proposals are of particular significance for
exploring novel phenomena of cavity QED and ultracold atoms.

This work is supported by NSFC under grants No. 11704175, the
NKBRSFC under grants No. 2011CB921500. X. C. Xie is supported in
part by US-DOE under Grant No. DE-FG02-04ER46124.


\begin{thebibliography}{99}
\bibitem{nondemolition} We note that, all the methods to
determine a BEC \cite{Gorlitz}, TG gas \cite{Paredes,Kinoshita}, BEC
to TG gas crossover \cite{Moritz}, or the sine-Gordon transition
\cite{Haller} of the 1D interacting bosonic gas are destructive.
\end{thebibliography}
\end{document}